# The low level radio frequency control system for DC-SRF photo-injector at Peking University[*]

WANG Fang(王芳)[1)] FENG Li-Wen(冯立文) LIN Lin(林林) HAO Jian-Kui(郝建奎) Quan Sheng-Wen(全胜文)
ZHANG Bao-Cheng(张保澄) LIU Ke-Xin(刘克新)

State Key Laboratory of Nuclear Physics and Technology, Peking University, Beijing 100871, China

**Abstract:** A low level radio frequency (LLRF) control system is designed and constructed at Peking University, which is for the DC-SRF photo injector operating at 2K. Besides with continuous wave (CW), the system is also reliable with pulsed RF and pulsed beam, the stability of amplitude and phase can achieve 0.13% and 0.1° respectively. It is worth noting that the system works perfectly when the cavity is driven at both generator driven resonator (GDR) and self-excited loop (SEL), the latter is useful in measuring the performance of the cavity.

Key words: LLRF, pulse mode, DC-SRF injector
PACS: 29.20.Ej

1 Introduction

Superconducting cavity technology has developed very rapidly over the past 25 years. While the first large scale applications at JLAB (CEBAF), KEK (TRISTAN), Argonne (ATLAS), CERN (LEP) are now in operation for many years, which are in CW operation. New projects under construction or in planning such as Euro-XFEL and ILC, have adopted superconducting cavities with pulse mode operation. Considerable experience of RF control at high gradients (>15 MV/m) with pulsed RF and pulsed beam has been gained at the TESLA Test Facility since 1998[1]. Several 500MHz superconducting cavities are adopted in the synchrotron light sources in China, and the sensitivity to Lorentz force detuning and microphonics is reduced in these cavities due to the high beam loading and the resulting low loaded Q factor. Peking University proposed a compact photocathode injector with medium beam current, DC-SRF injector. The upgraded injector consisting of a DC pierce gun and a 1.3GHz 3+1/2-cell superconducting cavity will be used in PKU-THz Facility and ERL-FEL as electron source.

Due to the high gradient and medium beam current, the bandwidth of the 3+1/2-cell cavity is narrow, then the cavity is susceptible to perturbations such as microphonics and Lorentz force detuning especially in the case of pulsed operation. In order to decrease the required RF power and the inconvenience during the operation of LLRF control system, the external quality factor Qe of the main coupler is set to $1\times 10^7$ and the corresponding bandwidth of the cavity is 130Hz. We studied on LLRF control system to maintain stable field in the injector since 2011 and now the system works in the operation of DC-SRF injector. The simulation results, main improvements and experiment results of the LLRF control system are presented in this paper.

[*] Supported by National Basic Research Project (2011CB808302)
1) E-mail: fangwang@pku.edu.cn

2 Simulation Results

The general controller used in LLRF control system is PI controller. To find the appropriate values for the controller, the behavior of the close loop is analyzed with the transfer function. If there is no detuning, the transfer function of the cavity is a first order system as a low-pass filter [2] and $\omega_{1/2}$ is 400Hz in our system. When detecting the amplitude or phase of the field, there is always delay with τ of 2μs typically [3] and detectors have limited bandwidth. The block diagram of the amplitude loop is illustrated in Fig. 1. It is hypothesized the bandwidth of the detector is very large and the transfer function is equal to 1. In this case, the loop is the simplest and optimized with gain margin of 10dB and phase margin of 60 ° for the open loop, with $k_p$=pm/3 and ki=w where pm=π/ (2τ$\omega_{1/2}$) and w=$\omega_{1/2}$. Fig.2 shows the bode plot and step response of the close loop, and indicates the system bandwidth is about 100 kHz while the setting time is about 10 μs.

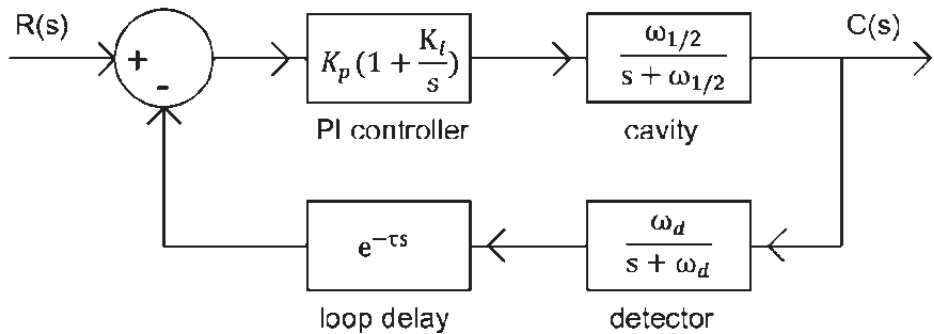

Fig.1: block diagram of the amplitude loop

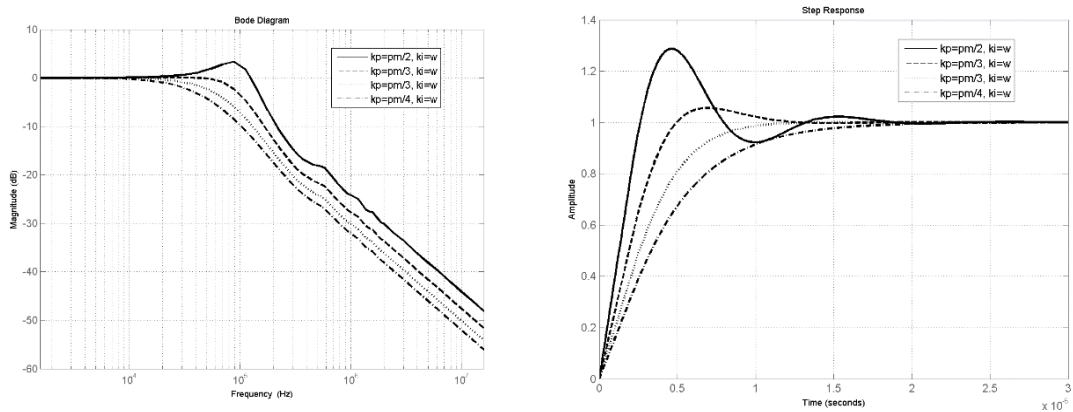

Fig.2: left)magnitude bode plot for the close loop, right)step response for the close loop

The behaviors of the close loop are also simulated besides the simplest case. Some useful conclusions are gained as below, when the detuning of the cavity is small, and the bandwidth of the detector is larger than the system bandwidth, there is no distinct effect on the loop behavior, otherwise the system bandwidth will be decreased and the setting time is increase, which should be avoided. Increasing the loop delay time decreases the system bandwidth and setting time. So the simulation gives directions of choosing controller parameters and RF elements in the whole system which should have enough bandwidth.

In our system, the bandwidth of the cavity is small and several sources of the field perturbations are changed rapidly, then the setting time is critical when designing the control system, so adjusting the PI controller is challenging.

3 Improvements of the LLRF control system

The LLRF control system consists of analog RF part and digital part as described previously [4]. The system works well when being tested at room temperature with bead pulling to disturb the field, but some problem encountered during the commissioning on the 3+1/2-cell cavity cooled to 2K. Switched from CW mode to pulse mode, the system can not maintain the field stable any more. It is finally found that the Lorentz force detuning during the filling stage causes large phase shift, which challenges the traditional PI controller due to its weakness about integrator-windup when the error is large. To solve this problem, a special controller with a predicative anti-windup PI strategy is designed to substitute the traditional one as illustrated in Fig.3, when the error is large than the threshold, only P works, otherwise PI works. With the new controller, the system works perfectly. During the improvement of the controller, the SEL mode is considered meanwhile, which is a good choice to measure the performance of 3+1/2-cell cavity as it follows changes in resonator frequency immediately and can therefore be used to establish gradient in the cavity with moderate RF power.

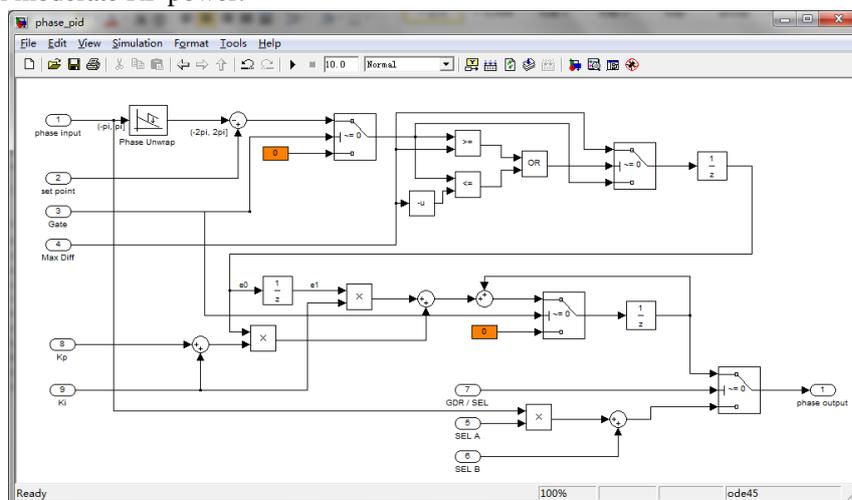

Fig. 3: Internal algorithm of phase controller in Simulink

Some other improvements also included after the first commissioning such as the timing system of the ADC and DAC, the clock of the ADC is from external clock but the clock of DAC is from FPGA to remove the chirp of the DAC output. A FIR filter is added besides the DC-block in FPGA. To speed data acquisition and signal process, GNU Radio and python are adopted to substitute Labview and the version of Human–Machine Interface is shown in Fig.4.

## 4 Experiment Results

After the upgraded of the controller, the LLRF control is reliable in both CW and pulse RF. When the 3+1/2-cell cavity is cooled to 2K in the case of RF pulse repetition of 10Hz and duty factor of 10%, the stable results were obtained with gradient up to 10MV/m, as illustrated in Fig.4 indicating the stability of amplitude and phase is 0.13% and 0.1° respectively. We gained stable electron beams with the current of 0.25 mA CW and 1.5mA at pulse mode [5], the energy of 3MeV at July of 2013 with the LLRF control system.

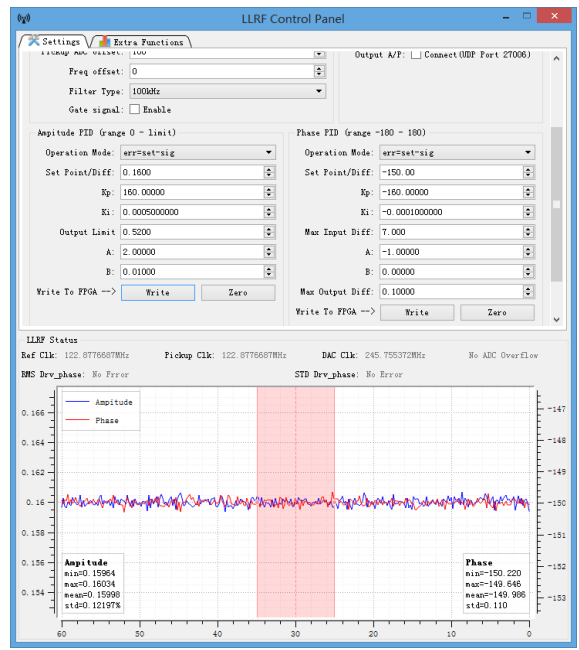

Fig.4: Long-term test of LLRF control for DC-SRF injector

Besides field stabilization the system is used to measure the major perturbation sources of the 3+1/2-cell cavity driven in SEL mode. With the helium pressure of 25.1mbar, the LLRF system tracks the resonant frequency of the cavity with different gradients in CW mode. The Lorentz force detuning has been measured range from 3.5MV/m to 11.6MV/m, and from 8MV/m to the higher, the radiation from the cavity is increased rapidly, it is concluded field emission-like happens. So only the lower gradients are adopted when fitting the Lorentz force detuning constant K, which is about 1.9Hz/(MV/m)$^2$ for the 3+1/2-cell cavity. With the gradient of 3.5MV/m, the helium pressure detuning is measured, and the constant is about 82Hz/mbar. The microphonics effect has also been investigated by utilizing the LLRF system. The frequency spectrum was measured as illustrated in Fig.5. The vibrations with the frequency of 2Hz and 14.5 Hz may affect the cavity.

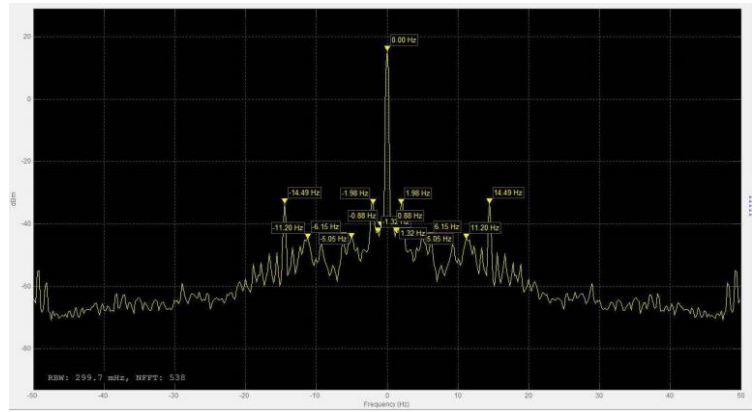

Fig.5: Real-time measurement of cavity microphonics

5 Conclusions

The LLRF system satisfies the requirements of the DC-SRF injector operation. Especially after the improvements of PI controller, the system works well in both CW and pulse RF, with the stability of amplitude and phase is 0.13% and 0.1° respectively. The system also works when cavity is driven in SEL mode which is useful in measuring the performance of the cavity itself. As the LLRF system is reliable and well understood in the DC-SRF injector, it is reproducible to other similar 1.3GHz superconducting cavities.